\documentclass[11pt,a4]{article}
\usepackage{latexsym,amssymb}

\newcommand{\be}{\begin{equation}}
\newcommand{\ee}{\end{equation}}
\newcommand{\beqs}{\begin{eqnarray}}
\newcommand{\eeqs}{\end{eqnarray}}

\def\({\left(}
\def\){\right)}

\newcommand{\rU}{\mathrm{U}}

\newcommand{\rSU}{\mathrm{SU}}

\newcommand{\rSO}{\mathrm{SO}}

\def\mxth{\mathsurround=0pt }
\def\xversim#1#2{\lower2.pt\vbox{\baselineskip0pt \lineskip-.5pt
x  \ialign{$\mxth#1\hfil##\hfil$\crcr#2\crcr\sim\crcr}}}

\def\IG{\relax{\rm I\kern-.18em \Gamma}}
\def\be{\begin{equation}}
\def\ee{\end{equation}}
\def\bea{\begin{eqnarray}}
\def\eea{\end{eqnarray}}

\def\bfone{\relax{\rm 1\kern-.35em 1}}

\begin{document}

\begin{titlepage}

\thispagestyle{empty}

\begin{flushright}
\hfill{CERN-PH-TH/2007-001}\\
\hfill{UCLA/07/TEP/1}
\end{flushright}

\vspace{35pt}

\begin{center}{ \LARGE{\bf
Critical points of the Black-Hole potential for homogeneous
special geometries}} \vspace{60pt}

{\bf  R. D'Auria$^\bigstar$, S. Ferrara$^\dag$ and M.
Trigiante$^\bigstar $}

\vspace{15pt}

$^\bigstar${\it Dipartimento di Fisica, Politecnico di Torino \\
C.so Duca degli Abruzzi, 24, I-10129 Torino, and\\
Istituto Nazionale di Fisica Nucleare, \\
Sezione di Torino,
Italy}\\[1mm] {E-mail: riccardo.dauria@polito.it,  mario.trigiante@polito.it}

$^\dag$ {\it CERN, Physics Department, CH 1211 Geneva 23,
Switzerland\\ and\\ INFN, Laboratori
Nazionali di Frascati, Italy\\and\\
Department of Physics \& Astronomy, University of California, Los Angeles, CA, USA}\\[1mm] {E-mail: Sergio.Ferrara@cern.ch}

\vspace{50pt}

{ABSTRACT}
\end{center}

\medskip

 We extend the analysis of N=2 extremal Black-Hole attractor equations to the case of special geometries based
 on homogeneous coset spaces. For non-BPS critical points
(with non vanishing central charge) the (Bekenstein-Hawking)
 entropy formula is the same as for symmetric spaces, namely four times the square of the central charge evaluated at the critical point.
For non homogeneous geometries the deviation from this formula is
given in terms of geometrical data of special geometry in presence
of a background symplectic charge vector.

\end{titlepage}

\newpage
\section{Introduction}
A remarkable feature of extremal  black-holes in supergravity is
the attractor mechanism which is at work for both BPS
\cite{attractor} and non-BPS \cite{fgk,fk,gijt,k,tt0,g,kss}
solutions: Starting from generic boundary values at radial
infinity, the scalar fields coupled to the black-hole evolve
towards fixed values at the horizon which are functions of the
quantized electric and magnetic charges only. These fixed values
are determined by the minimum of a certain effective potential
$V_{BH}$ which depends on the scalar fields and on the charges.
Extremal black holes in four dimensions feature a near-horizon
geometry of the form $AdS_2\times S^2$. The area of the horizon,
which is expressed in terms of the radius of $S^2$ in the throat
of the solution, is given by the value of $V_{BH}$ at its minimum
and is related via the Bekenstein-Hawking (B-H) formula, to the
entropy of the black hole
\begin{eqnarray}
S&=&\frac{A_H}{4}=\pi\,V_{BH\,|extr.}\,.\label{entropy}
\end{eqnarray}
 Although the
attractor mechanism was shown to be at work for BPS solutions also in the presence of higher
derivative corrections \cite{cdwkm}, we shall restrict our analysis to second order derivative
supergravity action. Extremal black holes play an important role in the
microscopic/statistical interpretation of the B-H expression for the entropy possibly
corrected by higher derivative contributions \cite{sv,msw,osv,d,ddmp,s,dst} (for a recent
review on black holes see \cite{p}).\par
 Recently the critical points of the black-hole
potential $V_{BH}$ in $N=2,\,D=4$ supergravity coupled to vector multiplets have been studied
and classified for all homogeneous symmetric special K\"ahler geometries \cite{bfgm}. It is
the goal of the present paper to extend the analysis to homogeneous non-symmetric spaces whose
general classification was given and discussed in \cite{a,dwvvp,c}. The nature of these spaces
is quite interesting from a physical point of view since some of them naturally appear in
brane dynamics, when the brane and bulk degrees of freedom are unified in the four dimensional
effective supergravity. Homogeneous special geometries can be classified by coset spaces
$\mathcal{M}=G/H$ where
\begin{eqnarray}
G&=&\left[\rSO(1,1)_0\times \rSO(2,2+q)\right]\ltimes {\mathcal
N}\,,\nonumber\\
 H&=&\rSO(2)\times \rSO(2+q)\,,
\end{eqnarray}
and ${\mathcal N}$ is a subgroup generated by a nilpotent graded
subalgebra $\mathbb{N}$ of the isometry algebra:
\begin{eqnarray}
\mathcal{N}&=&\exp{(\mathbb{N})}\,,\nonumber\\
\mathbb{N}&=&W^{(+1)}\oplus W^{(+2)}\,,
\end{eqnarray}
where the superscripts refer to the $\rSO(1,1)_0$--grading. The
generators $T_\alpha$ of $W^{(+1)}$ transform in a real (in
general reducible) $d$-dimensional spinorial representation of
$\rSO(2,q+2)$ while $W^{(+2)}$ is one--dimensional and its the
generator $T_\bullet$ is an $\rSO(2,q+2)$--singlet. The algebraic
structure of $\mathbb{N}$ is defined by the following commutation
relations:
\begin{eqnarray}
\left[T_{\alpha},\,T_{\beta}\right]&=&2\,C_{\alpha\beta}\,T_{\bullet}\,\,;\,\,\,\left[T_{\alpha},\,T_\bullet\right]=0\,,
\end{eqnarray}
$C_{\alpha\beta}$ being the antisymmetric real matrix to be
identified with the charge conjugation matrix of the spinorial
representation defined by $T_\alpha$ (see appendix \ref{spinor}).
The number of complex coordinate of the manifold is
\begin{eqnarray}
n&=&\mbox{dim}_{\mathbb{C}}\mathcal{M}=3+q+\frac{d}{2}\,.
\end{eqnarray}
Denoting by $T_{\Lambda\Sigma}=-T_{\Sigma\Lambda}$
($\Lambda,\,\Sigma=0,1,\dots, q+3$) the $\rSO(2,2+q)$ generators
and by $h_0$ the $\rSO(1,1)_0$ generator, the following
commutation relations hold:
\begin{eqnarray}
\left[T_{\Lambda\Sigma},\,T_{\Gamma\Delta}\right]&=&\frac{1}{2}\left(\eta_{\Lambda\Delta}\,T_{\Sigma\Gamma}+\eta_{\Sigma\Gamma}\,T_{\Lambda\Delta}-
\eta_{\Lambda\Gamma}\,T_{\Sigma\Delta}-\eta_{\Sigma\Delta}\,T_{\Lambda\Gamma}\right)\,,\nonumber\\
\left[T_{\Lambda\Sigma},\,T_{\alpha}\right]&=&-\frac{1}{4}\,(\Gamma_{\Lambda\Sigma})_\alpha{}^\beta\,T_\beta\,\,;\,\,\,
\left[h_{0},\,T_{\alpha}\right]=
T_{\alpha}\,\,;\,\,\,\left[h_{0},\,T_{\bullet}\right]=
2\,T_{\bullet}\,,
\end{eqnarray}
all the other commutators between the same generators being zero.
The $d/2$ complex coordinates parametrizing $T_\alpha$ define an
$\rSO(2,2+q)$ Clifford module. The compact isometries are
$\rSO(2)\times\rSO(2+q)\times \mathcal{S}_q(P,\dot{P})$, and
include the centralizer $\mathcal{S}_q(P,\dot{P})$ of the Clifford
algebra representation \cite{dwvvp}, which is not contained inside
$G$. In the table below we summarize the notations introduced in
\cite{dwvvp} for the classification of homogeneous special
K\"ahler manifolds.
\begin{center}
\begin{tabular}{|c|c|}\hline
$\mathcal{M}$  & $d$ \\\hline
  $L(0,P,\dot{P})$ & $2\,(P+\dot{P})$ \\\hline
  $L(q,P,\dot{P}),\,q=4\,m>0$ & $4\,D_q\,(P+\dot{P})$ \\\hline
  $L(q,P), q>0, q\neq 4\,m$& $4\,D_q\,P$ \\\hline
  $L(-1,P)$ & $2\,P$ \\\hline
  $L(-2,P)$ & $2\,P$ \\ \hline
\end{tabular}
\end{center}\label{tab1}
where the values for $D_q$ are:
\begin{eqnarray}
&&D_1=1\,\,,\,\,\,D_2=2\,\,,\,\,\,D_3=D_4=4\,\,,\,\,\,D_5=D_6=D_7=D_8=8\,.
\end{eqnarray}
Certain  homogeneous (non--symmetric) spaces listed in Table
\ref{tab1} naturally appear in brane dynamics, for instance in
Type IIB compactification on $K3\times T^2/\mathbb{Z}_2$
orientifold in the presence of space--filling $D3$ and $D7$ branes
\cite{tt,aft,adft}. In this case the relevant space is
$L(0,P,\dot{P})$, which is parametrized by the $S,T,U$ moduli of
the bulk as well as the $P$ and $\dot{P}$ (which may be different
from $P$) complex moduli describing the positions of the $D3$ and
$D7$ branes along $T^2$ respectively. The model has the additional
global symmetry $\mathcal{S}_0(P,\dot{P})=\rSO(P)\times
\rSO(\dot{P})$ which is not part of the isometry group $G$, whose
compact subgroup is $\rSO(2)\times \rSO(2)$ in this case. This
state of affair is actually a particular case of a more general
feature of $L(q,P,\dot{P})$ spaces for which the rigid symmetry
$S_q(P,\dot{P})$ depends in general on $q$ as well as on the
number of Clifford moduli.\par The (rank three) symmetric spaces
fall in the class $L(0,P)$ (which is the same as $L(q,0)$ for
$q=P$) and $L(q,1)$ with $q=1,2,4,8$ which correspond to the
\emph{magic} supergravities. The spaces $L(q,0)$ are obtained by
setting $T_\alpha=0$ and define the infinite series of manifolds
of the form
\begin{eqnarray}
\mathcal{M}&=&\frac{\rSU(1,1)}{\rU(1)}\times\frac{\rSO(2,2+q)}{\rSO(2)\times
\rSO(q+2)}\,.
\end{eqnarray}
In this context the STU model corresponds to $L(0,0)$ and is the
smallest rank three symmetric space. In the magic supergravity
cases the real dimension of the Clifford module is respectively
given by: $d=4,\,8,\,16,\,32$, corresponding to the
\emph{Freudenthal triple system} of the Jordan algebras on
$\mathbb{R},\,\mathbb{C},\,\mathbb{H},\,\mathbb{O}$ respectively.
 For
all the symmetric spaces the $H$--connection $\Omega^{(H)}$
actually coincides with the torsion--free Riemann connection
$\omega$, while they differ in the non--symmetric case. This will
be explicitly shown in the next section.\par The black-hole
attractors for the $S,T,U$ model are clearly connected to the
$N=8$ attractors studied in \cite{fk}. On the other hand the black
hole attractors for all symmetric spaces were studied in
\cite{bfgm}. There it was found that such attractors fall into
three classes of orbits: One BPS and two non-BPS. The latter were
further distinguished by the fact that the $N=2$ central charge
$Z$ is non-vanishing in one case and vanishing in the other. It is
the scope of the present paper to generalize this classification
to the homogeneous spaces which are not symmetric.\par We actually
find that for the non-BPS orbit with $|Z|\neq 0$ the same rule as
for the symmetric case applies, being
\begin{eqnarray}
V_{BH\,\left| extr.\right.}&=&4\,|Z|^2_{\left| extr.\right.}\,.
\end{eqnarray}
The paper is organized as follows. In section \ref{sec2} we give
the special geometry for homogeneous spaces in rigid, namely
tangent space, indices. We further compute the Riemann spin
connection and the H--connection for the cosets. In sections
\ref{sec3} and \ref{sec4} we analyze the attractor equations for
both $Z\neq 0$ and $Z=0$ orbits. While in the case $Z\neq 0$ the
situation is similar to the symmetric case (in particular we get
the same expression for the entropy in terms of the central
charge), the classification is more involved in the $Z=0$ case,
where many orbits seem to exist. Finally some useful mathematical
tools for special geometry describing homogeneous spaces are
collected in the appendices.
\section{Special geometry in rigid indices}\label{sec2}
In the present section we shall explicitly construct the tensor
quantities on a homogeneous special K\"ahler manifold, which will
be relevant to our discussion in a rigid basis. In appendix
\ref{ska} we review some basic facts and definitions about special
geometry. The following discussion however relies only on the
isometry properties of these spaces, outlined in the introduction,
and will not make use of special K\"ahler identities.
\par We start from writing the connection and curvature in a real
basis of the tangent space of the manifold and then introduce the
rigid metric and complex structure. We shall choose as a basis for
the tangent space of the symmetric submanifold
$\rSO(2,2+q)/\rSO(2)\times\rSO(q+2)$ the $2\,(2+q)$ non-compact
generators $T_{aI}$, $a=0,1$ and $I=2,\dots, q+3$, defined by the
Cartan decomposition of the $\mathfrak{so}(2,2+q)$ algebra. The
corresponding basis of the tangent space of the whole homogeneous
manifold will therefore consist of the following generators
\begin{eqnarray}
\{T_A\}&=&\{h_0,\,T_{aI},\,T_\alpha,\,T_\bullet\}\,\,\,;\,\,\,\,A=1,\dots,
2n\,.
\end{eqnarray}
 The
vielbeins dual to the isometry generators $\{T_A\}$ are
$\{V^A\}=\{V^0,\,V^{aI},\,V^\alpha,\,V^\bullet\}$ which, together
with the $\rSO(2)\times\rSO(q+2)$--connection 1--forms
$\Omega^{(H)\,\lambda}=(\Omega^{(H)\,ab},\,\Omega^{(H)\,IJ})$,
satisfy the following set of Maurer--Cartan equations:
\begin{eqnarray}
dV^0&=&0\,,\nonumber\\
dV^{aI}+\frac{1}{2}\,\Omega^{(H)\,a}{}_b\wedge
V^{bI}+\frac{1}{2}\,\Omega^{(H)\,I}{}_J\wedge V^{aJ}&=&0\,,\nonumber\\
dV^{\alpha}-\frac{1}{8}\,(\Gamma_{ab})_\beta{}^\alpha\,\Omega^{(H)\,ab}\wedge
V^\beta
-\frac{1}{8}\,(\Gamma_{IJ})_\beta{}^\alpha\,\Omega^{(H)\,IJ}\wedge
V^\beta+V^0\wedge V^\alpha&=&0\,,\nonumber\\
dV^\bullet+C_{\alpha\beta}\,V^\alpha\wedge V^\beta+2V^0\wedge V^\bullet&=&0\,.\label{mc}
\end{eqnarray}
The metric $g_{AB}$ on the tangent space at the origin is defined
as follows:
\begin{eqnarray}
g_{AB}&:&\cases{g_{\bullet\bullet}=1=g_{00}\cr g_{aI,
bJ}=\frac{1}{8}\,\delta_{ab}\,\delta_{IJ}\cr g_{\alpha\beta}=\delta_{\alpha\beta}}\,.
\end{eqnarray}
The metric connection $\omega_A{}^B$ satisfies the property
\begin{eqnarray}
\omega_A{}^C\,g_{CB}&=&-\omega_B{}^C\,g_{CA}\,,
\end{eqnarray}
and is the solution of the zero--torsion condition:
\begin{eqnarray}
T^A&\equiv& DV^A=dV^A+\omega_B{}^A\wedge V^B=0\,.
\end{eqnarray}
The explicit form for the spin connection is:
\begin{eqnarray}
\omega_{B}{}^C&=&\tilde{\omega}_{B}{}^C+\omega^{(H)}{}_{B}{}^C\,,\nonumber\\
\omega^{(H)}{}_{B}{}^C&=&\Omega^{(H)\,\lambda}\,C_{\lambda
B}{}^C\,,\nonumber\\
\tilde{\omega}_{B}{}^C&=&\frac{1}{2}\,\left(C_{AB}{}^C+g^{CC^\prime}\,g_{AA^\prime}\,C_{C^\prime
B }{}^{A^\prime}+g^{CC^\prime}\,g_{BB^\prime}\,C_{C^\prime A
}{}^{B^\prime}\right)\,V^A\,,\label{omomt}
\end{eqnarray}
where $\omega^{(H)}{}_{B}{}^C$ is the $H$--connection which is
expressed in terms of the $\Omega^{(H)\,\lambda}$--forms,
determined by solving the second of eqs. (\ref{mc}), see eqs.
(\ref{omH}). In the symmetric case the isometry group $G$ is
generated by a semisimple Lie algebra $\mathfrak{g}$ which can be
decomposed into its maximal compact subalgebra
$\mathfrak{h}\subset\mathfrak{g}$ and the orthogonal complement
$\mathfrak{t}$ to $\mathfrak{h}$ which is spanned by non--compact
generators (Cartan decomposition). Choosing the basis $\{T_A\}$
for the tangent space of the manifold in $\mathfrak{t}$, we have
$C_{AB}{}^C\equiv 0$, which in turn implies $\tilde{\omega}\equiv
0$ in eqs. (\ref{omomt}). Therefore in the symmetric case we can
use the Cartan decomposition of the isometry algebra to define a
basis of generators with respect to which the spin connection
coincides with the $H$--connection: $\omega\equiv\omega^{(H)}$.
This is not the case for non--symmetric manifolds, for which
$\tilde{\omega}$ is always non--vanishing. The components of
$\tilde{\omega}$ are given in Appendix \ref{omegaR}.\par
 In terms of $\omega$ the curvature 2--form is defined as:
\begin{eqnarray}
R_A{}^B&=&d\omega_A{}^B-\omega_A{}^C\wedge
\omega_C{}^B=-\frac{1}{2}\,R_{CD,A}{}^B\,V^C\wedge V^D\,,
\end{eqnarray}
or in components:
\begin{eqnarray}
R_{AB,C}{}^D&=&-2\,\tilde{\omega}_{[B,\vert
C}{}^E\,\tilde{\omega}_{A],E}{}^D+C_{AB}{}^\lambda\,C_{\lambda,C}{}^D+C_{AB}{}^E\,\tilde{\omega}_{E,C}{}^D\,.\label{rcomp}
\end{eqnarray}
 The explicit form of the components of the
Riemann tensor in the chosen rigid basis is given in eq.
(\ref{Rcomponent}) of Appendix \ref{omegaR}.\par
 Let us define on the tangent
space the following complex structure:
\begin{eqnarray}
V^S&=&\frac{1}{\sqrt{2}}\,(V^\bullet-i\,V^0)\,\,;\,\,\,V^I=\frac{1}{\sqrt{2}}\,(V^{1I}+i\,V^{0I})\,,\nonumber\\
V^i&=&\frac{1}{\sqrt{2}}\,(V^{i_1}+i\,V^{i_2})\,\,\,,\,\,\,\,\alpha=(i_1,\,i_2),\,\,\,i_1,i_2=1,\dots,\frac{d}{2}\,,
\end{eqnarray}
where in the indices $i_1$ and $i_2$ the charge conjugation matrix
reads:
\begin{eqnarray}
C_{i_1i_2}&=&\delta_{i_1i_2}=-C_{i_2i_1}\,.
\end{eqnarray}
We shall use the indices $r,\,s,\dots$ to label the complex
vielbein basis: $V^r=(V^S,\,V^I,\,V^i)$. In this basis the rigid
metric has the following form:
\begin{eqnarray}
g_{S\bar{S}}&=&1\,\,;\,\,\,g_{I\bar{J}}=\frac{1}{8}\,\delta_{I\bar{J}}\,\,;\,\,\,g_{i\bar{\jmath}}=\delta_{i\bar{\jmath}}\,,
\end{eqnarray}
and the  curvature tensor reads:
\begin{eqnarray}
R_p{}^q&=&R_{r\,\bar{s},\,p}{}^q\,V^r\wedge
\overline{V}^{\,\,
\bar{s}}\,,\nonumber\\
R_{r\,\bar{s},\,p}{}^q&=&g_{r\bar{s}}\,\delta_p^q+g_{\bar{s}p}\,\delta_r^q-C_{rpk}\,\bar{C}_{\bar{s}}{}^{kq}\,,
\end{eqnarray}
where the non vanishing components of the symmetric tensor
$C_{rsp}$ are:
\begin{eqnarray}
C_{SIJ}&=&\frac{1}{8}\,\delta_{IJ}\,\,\,;\,\,\,\,C_{Iij}=\frac{1}{4}\,({\bf
\Gamma}_I)_{ij}\,,
\end{eqnarray}
and the (complex) gamma matrices $({\bf \Gamma}_I)_{ij}$ are
defined in (\ref{complexcliff0}).
 The $C_{rsp}$ is manifestly $\rSO(q+2)$--invariant and satisfies the relation:
\begin{eqnarray}
C_{rsp}\,\overline{C\,}^p{}_{(\bar{r}_1\bar{r}_2}\,\overline{C\,}^s{}_{\bar{r}_3\bar{r}_4)}-\frac{4}{3}\,g_{r(\bar{r}_1}\,\overline{C}_{\bar{r}_2\bar{r}_3\bar{r}_4)}&=&
E_{r\,\bar{r}_1\bar{r}_2\bar{r}_3\bar{r}_4}\,,\label{Etensor}
\end{eqnarray}
where the non--vanishing components of the tensor $E$ are:
\begin{eqnarray}
E_{S\,\overline{\imath\jmath
kl}}&=&\frac{1}{2}\,\Gamma_{\overline{\imath\jmath
kl}}\,\,;\,\,\,E_{i\,\bar{I}\overline{\jmath
kl}}=-\frac{1}{8}\,({\bf
\Gamma}_I)_i{}^{\bar{\imath}}\,\Gamma_{\overline{\imath\jmath
kl}}\,,\label{E}\\
\Gamma_{ijkl}&=&\sum_{I=2}^{q+3}\,({\bf \Gamma}_I)_{(ij}\,({\bf
\Gamma}_I)_{kl)}=\sum_{m=3}^{q+3}\,(\Gamma_m)_{(ij}\,(\Gamma_m)_{kl)}-\delta_{(ij}\,\delta_{kl)}\,,
\end{eqnarray}
In the symmetric case $\Gamma_{ijkl}=0$ and therefore also the
tensor $E$ vanishes identically.
\section{The attractor equations and  non--BPS solutions with $Z\neq
0$}\label{sec3} Consider the SUSY and matter central charges, the
latter written in rigid indices: $Z,\, Z_r=(Z_S,\,Z_I,\,Z_i)$. In
the $N=2$ theory the matter charges $Z_r$ are expressed as
derivatives of the SUSY central charge $Z$: $Z_r=D_r Z$. In terms
of these charges the effective potential $V_{BH}$ has the
following expression
\begin{eqnarray}
V_{BH}&=&|Z|^2+g^{r\bar{r}}\,Z_r\overline{Z}_{\bar{r}}\,.
\end{eqnarray}
Since the attractor point is defined by minimizing $V_{BH}$, the
attractor equations are $D_r V_{BH}=0$, which, by using the
identity (\ref{sk1}) of special geometry,
 can be recast in the form
\begin{eqnarray}
2\,\overline{Z}\,Z_r+i\,C_{rsp}\,g^{s\bar{s}}\,g^{p\bar{p}}\,\overline{Z}_{\bar{s}}\,\overline{Z}_{\bar{p}}&=&0\,.\label{attractor}
\end{eqnarray}
Equations (\ref{attractor}) can then be grouped in the following
way:
\begin{eqnarray}
(a)&:&2\,\overline{Z}\,Z_S+8\,i\,\sum_{I}\,(\overline{Z}_{{\bar I}})^2=0\,,\nonumber\\
(b)_I&:&2\,\overline{Z}\,Z_I+2\,i\,\overline{Z}_{\bar{S}}\,\overline{Z}_{\bar{I}}+\frac{i}{4}\,({\bf
\Gamma}_I)_{ij}\,\overline{Z}_{\bar{\imath}}\,\overline{Z}_{\bar{\jmath}}=0\,,\nonumber\\
(c)_i&:&2\,\overline{Z}\,Z_i+4\,i\,({\bf
\Gamma}_I)_{ij}\,\overline{Z}_{\bar{I}}\,\overline{Z}_{\bar{\jmath}}=0\,.
\end{eqnarray}
From equations $(c)_i$ and the property (\ref{complexcliff}) we
deduce, in the case $Z\neq 0$, the following relations:
\begin{eqnarray}
({\bf
\Gamma}_I)_{ij}\,\overline{Z}_{\bar{I}}\,\overline{Z}_{\bar{\jmath}}&=&
\frac{i}{2}\,\,\overline{Z}\,Z_i\,,\label{user1}\\
({\bf
\Gamma}_I)_{ij}\,{Z}_{I}\,\overline{Z}_{\bar{\jmath}}&=&\frac{2\,i}{Z}\,Z_i\,\sum_I\,(Z_I)^2\,.\label{user2}
\end{eqnarray}
Consider now the following combination of equations $(b)_I$:
\begin{eqnarray}
(b)_I\,({\bf \Gamma}_I)_{ij}\,\overline{Z}_{\bar{\jmath}}&:&0=
2\,({\bf
\Gamma}_I)_{ij}\,\overline{Z}\,Z_I\,\overline{Z}_{\bar{\jmath}}+2\,i\,\overline{Z}_{\bar{S}}\,({\bf
\Gamma}_I)_{ij}\,\overline{Z}_{\bar{I}}\,\overline{Z}_{\bar{\jmath}}+\frac{i}{4}\,({\bf
\Gamma}_I)_{ij}({\bf
\Gamma}_I)_{kl}\,\,\overline{Z}_{\bar{\jmath}}\,\,\overline{Z}_{\bar{k}}\,\,\overline{Z}_{\bar{l}}=\nonumber\\
&&=4\,i\,\frac{\overline{Z}}{Z}\,Z_i\,\sum_I\,(Z_I)^2-\overline{Z}_{\bar{S}}\,\overline{Z}\,Z_i+\frac{i}{4}\,\Gamma_{ijkl}\,\overline{Z}_{\bar{\jmath}}\,\,\overline{Z}_{\bar{k}}\,\,\overline{Z}_{\bar{l}}\,,\label{bi}
\end{eqnarray}
where we have used the relations (\ref{user1}) and (\ref{user2}).
Equation $(a)$ can then be written in the form
\begin{eqnarray}
\sum_I\,(Z_I)^2&=&-\frac{i}{4}\,Z\,\overline{Z}_{\bar{S}}\,.\label{zs}
\end{eqnarray}
Using (\ref{zs}), equation (\ref{bi}) yields
\begin{eqnarray}
\Gamma_{ijkl}\,\overline{Z}_{\bar{\jmath}}\,\,\overline{Z}_{\bar{k}}\,\,\overline{Z}_{\bar{l}}&=&0\,,
\end{eqnarray}
which, in virtue of eqs. (\ref{E}), implies that:
\begin{eqnarray}
\overline{E}_{\bar{r}\,r_1r_2r_3r_4}\,\overline{Z}^{\,\,r_1}\,\overline{Z}^{\,\,\,\,
r_2}\,\overline{Z}^{\,\,r_3}\,\overline{Z}^{\,\,r_4}&=&0\,.
\end{eqnarray}
Note that this result holds also in the $Z=0$ case. On the other
hand, from eqs. (\ref{attractor}) and (\ref{Etensor}) we find the
relation:
\begin{eqnarray}
\overline{Z}\,\overline{Z}_{\bar{r}}\,\left(|Z|^2-\frac{1}{3}\,Z_r\,\overline{Z}_{\bar{s}}\,g^{r\bar{s}}\right)&=&-\frac{i}{8}\,
\,\overline{E}_{\bar{r}\,r_1r_2r_3r_4}\,\overline{Z}^{\,\,r_1}\,\overline{Z}^{\,\,\,\,
r_2}\,\overline{Z}^{\,\,r_3}\,\overline{Z}^{\,\,r_4}\,,
\end{eqnarray}
from which we deduce that for all solutions to eq.
(\ref{attractor}) with $Z\neq 0$ and $\overline{Z}_{\bar{r}}\neq
0$ (non--BPS) the following sum-rule holds\footnote{This sum-rule
was also found in \cite{tt0} for a class of non-BPS solutions of
special geometries based on generic cubic prepotentials
(d-geometries).}:
\begin{eqnarray}
|Z|^2+Z_r\,\overline{Z}_{\bar{s}}\,g^{r\bar{s}}&=&4\,|Z|^2\,,
\end{eqnarray}
namely that the  non--BPS attractors with $Z\neq 0$ feature the
same value of the invariant which characterizes the corresponding
orbit in the symmetric case and which is related to the entropy by
eq (\ref{entropy})
\begin{eqnarray}
S&=&4\pi\,|Z|^2_{|extr.}\,.
\end{eqnarray}
 For BPS solutions $Z_r\equiv 0$ and the entropy reduces to
\begin{eqnarray}
S&=&\pi\,|Z|^2_{|extr.}\,.
\end{eqnarray}
\section{The  $Z=0$ non-BPS attractors}\label{sec4}
If $Z=0$ the attractor equations in rigid indices
(\ref{attractor}) reduce to
\begin{eqnarray}
C_{rsp}\,g^{s\bar{s}}\,g^{p\bar{p}}\,\overline{Z}_{\bar{s}}\,\overline{Z}_{\bar{p}}&=&0\,,\label{attractor0}
\end{eqnarray}
which can also be written as follows
\begin{eqnarray}
(a)&:&\sum_{I}\,(\overline{Z}_{{\bar I}})^2=0\,,\nonumber\\
(b)_I&:&\overline{Z}_{\bar{S}}\,\overline{Z}_{\bar{I}}+\frac{1}{8}\,({\bf
\Gamma}_I)_{ij}\,\overline{Z}_{\bar{\imath}}\,\overline{Z}_{\bar{\jmath}}=0\,,\nonumber\\
(c)_i&:&({\bf
\Gamma}_I)_{ij}\,\overline{Z}_{\bar{I}}\,\overline{Z}_{\bar{\jmath}}=0\,.\label{attractor02}
\end{eqnarray}
Recall that $Z_i,\,\overline{Z}_{\bar \imath}$ are in general
$P+\dot{P}$ copies of the $\rSO(q+2)$ spinorial representation and
there is a further $\mathcal{S}_q(P,\,\dot{P})$ global symmetry
mixing the various copies. The compact global symmetry group of
the theory is therefore $\rSO(2)\times\rSO(2+q)\times
\mathcal{S}_q(P,\,\dot{P})$.\par
 Let us
first give a general discussion of these attractors and then
consider some explicit cases. \paragraph{$Z_I\neq 0$ case.}
Consider first the case in which $Z_I$ are not all vanishing.
Equation $(a)$ can be solved by taking as only non-vanishing
charges $Z_I$, $Z_2$ and $Z_3$ and further requiring that
$Z_2=-i\epsilon Z_3$, where $\epsilon=\pm 1$. The group
$\rSO(q+2)$ is now broken to $\rSO(q)$. Equations $(c)_i$ impose
then a ``chirality'' constraint on the spinorial charges $Z_i$
\begin{eqnarray}
(\gamma_3)_{ij}\,\overline{Z}_{\bar{\jmath}}&=&\epsilon\,\overline{Z}_{\bar{\imath}}\,,\label{chiral}
\end{eqnarray}
while equations $(b)_I$ can be solved in $Z_S$ to give
\begin{eqnarray}
\overline{Z}_{\bar{S}}&=&-\frac{\epsilon}{8\,\overline{Z}_3}\,\sum_i(\overline{Z}_{\bar{\imath}})^2\,.\label{zzs}
\end{eqnarray}
If the non-vanishing $Z_i,\,\overline{Z}_{\bar{\imath}}$ consist
of just one copy of the spinorial representation, the $\rSO(q)$ is
further broken to the stability group of the $\rSO(q)$ spinorial
representations and $\mathcal{S}_q(P,\,\dot{P})$ is broken either
to $\mathcal{S}_q(P-1,\,\dot{P})$ or to
$\mathcal{S}_q(P,\,\dot{P}-1)$. If more independent copies of the
spinorial representation are non-vanishing then the stability
group will in general be smaller.
\paragraph{$Z_I= 0$ case.} The symmetry group $\rSO(q+2)$ is only broken
by the spinorial charges to their stability subgroup. The only non
trivial equations are the $(b)_I$ which read
\begin{eqnarray}
({\bf
\Gamma}_I)_{ij}\,\overline{Z}_{\bar{\imath}}\,\overline{Z}_{\bar{\jmath}}=0\,.\label{ps}
\end{eqnarray}
These equations have in general solutions if more copies of the
spinorial representation are present, namely if $(P+\dot{P})>1$.
For $q>1$, we can also have solutions of (\ref{ps}) if
$P+\dot{P}=1$. The general solution may  therefore involve one or
more  copies of the spinorial representation,  thus breaking
$\rSO(q+2)$ to the stabilizer of the spinorial representation or
to a smaller group respectively. Below we shall work out three
explicit examples of non symmetric manifolds.
\subsection{The  $L(0,P,\dot{P})$ space}

If $q=0$ the spinor charges $Z_i$ split into two sets which we
shall denote by $\chi_m$, $\chi_{\dot{m}}$, where $m=1,\dots, P$
and $\dot{m}=1,\dots, \dot{P}$. The compact global symmetry group
of the theory is $\rSO(2)\times \rSO(2)\times \rSO(P)\times
\rSO(\dot{P})$. The attractor equations (\ref{attractor02}) read:
\begin{eqnarray}
(a)&:&(\overline{Z}_{\bar{2}})^2+(\overline{Z}_{\bar{3}})^2=0\,,\nonumber\\
(b)_2&:&\overline{Z}_{\bar{S}}\,\overline{Z}_{\bar{2}}+\frac{i}{8}\,(\sum_{m}\,(\overline{\chi}_{{\bar
m}})^2 + \sum_{\dot{m}}\,(\overline{\chi}_{{\dot{ \bar{m}}}})^2)=0\,,\nonumber\\
(b)_3&:&\overline{Z}_{\bar{S}}\,\overline{Z}_{\bar{3}}+\frac{1}{8}\,(\sum_{m}\,(\overline{\chi}_{{\bar
m}})^2 - \sum_{\dot{m}}\,(\overline{\chi}_{{\dot{\bar{m}}}})^2)=0\,,\nonumber\\
(c)_m&:&(\overline{Z}_{\bar{2}}-i\,\overline{Z}_{\bar{3}})\,\overline{\chi}_{{\bar
m}}=0\,,\nonumber\\
(c)_{\dot{m}}&:&(\overline{Z}_{\bar{2}}+i\,\overline{Z}_{\bar{3}})\,\overline{\chi}_{\dot
{\bar{m}}}=0 \,.\label{attractorL0PP}
\end{eqnarray}
Let us consider the various orbits of solutions.
\paragraph{The $Z_S\neq 0,\,Z_I\neq 0$ case.} If
$\overline{Z}_{\bar{2}}=i\,\overline{Z}_{\bar{3}}\neq 0$ equation
$(a)$ and equations $(c)_m$ are satisfied while equations
$(c)_{\dot{m}}$ imply $\chi_{\dot{m}}=0$. Equations $(b)_I$ imply
\begin{eqnarray}
\overline{Z}_{\bar{S}}&=&-\frac{i}{8\,\overline{Z}_{\bar{2}}}\,\sum_{m}\,(\overline{\chi}_{{\bar
m}})^2\,.
\end{eqnarray}
These solutions define an orbit with stability group
$\rSO(P-1)\times \rSO(\dot{P})$.
\paragraph{The $Z_S= 0,\,Z_I\neq 0$ case.} If
$\overline{Z}_{\bar{2}}=i\,\overline{Z}_{\bar{3}}\neq 0$ we still
have $\chi_{\dot{m}}=0$ while from equations $(b)_I$ we have
$\sum_{m}\,(\overline{\chi}_{{\bar m}})^2=0$, which may have a
trivial solution $\overline{\chi}_{{\bar m}}=0$ with stability
group $\rSO(P)\times \rSO(\dot{P})$ and a non--trivial solution
\begin{eqnarray}
\overline{\chi}_{{\bar 1}}&=&\pm i\,\overline{\chi}_{{\bar 2}}\neq
 0\,\,\,\,\,\overline{\chi}_{{\bar
m}\neq 1,2}=0\,,
\end{eqnarray}
with stability group $\rSO(P-2)\times \rSO(\dot{P})$.
\paragraph{The $Z_S\neq 0,\,Z_I=0$ case.} Equations $(b)_I$ imply
\begin{eqnarray}
\sum_{m}\,(\overline{\chi}_{{\bar
m}})^2&=&\sum_{\dot{m}}\,(\overline{\chi}_{{\dot{\bar{m}}}})^2=0\,,
\end{eqnarray}
which can have the following solutions with the corresponding
stability groups
\begin{eqnarray}
\overline{\chi}_{{\bar
m}}=\overline{\chi}_{{\dot{\bar{m}}}}=0\,\,\,&&\mbox{stability
group:}\,\,\rSO(2)\times \rSO(P)\times \rSO(\dot{P})\,,\nonumber\\
\left.\matrix{\overline{\chi}_{{\bar 1}}=\pm
i\,\overline{\chi}_{{\bar 2}}\neq 0\cr\overline{\chi}_{{\bar
m}\neq
1,2}=0=\overline{\chi}_{{\dot{\bar{m}}}}}\right\}\,\,\,&&\mbox{stability
group:}\,\,\rSO(P-2)\times \rSO(\dot{P})\,,\nonumber\\
\left.\matrix{\overline{\chi}_{{\bar 1}}=\pm
i\,\overline{\chi}_{{\bar 2}}\neq 0 & \overline{\chi}_{{\bar
m}\neq 1,2}=0\cr \overline{\chi}_{{\dot{\bar{1}}}}=\pm
i\,\overline{\chi}_{{\dot{\bar{2}}}}
&\overline{\chi}_{{\dot{\bar{m}}\neq
1,2}}=0}\right\}\,\,\,&&\mbox{stability
group:}\,\,\rSO(P-2)\times \rSO(\dot{P}-2)\,,\nonumber\\
\end{eqnarray}
\subsection{The $L(1,2)$ space}
The global symmetry group in this case is
$\rSO(2)\times\rSO(3)\times \rSO(2)$, $\rSO(2)$ being
$\mathcal{S}_1(2)$ for this manifold. Here we shall denote by
$\vec{\chi}_m\equiv (\chi_{\alpha,m})$ the spinorial central
charges $Z_i\equiv Z_{\alpha,m}$, where $\alpha=1,2$ is the
spinorial index while $m=1,2$ runs over the number of copies. The
vector charges are $Z_I=(Z_2,Z_3,Z_4)$. We can choose the relevant
gamma matrices to have the form:
\begin{eqnarray}
{\bf \Gamma}_2&=&i\,\bfone_2\times\bfone_2\,\,;\,\,\,{\bf
\Gamma}_3=\sigma_3\times \bfone_2\,\,;\,\,\,{\bf
\Gamma}_4=\sigma_1\times \bfone_2\,.
\end{eqnarray}
If we choose $Z_4=0$ and $Z_2=-i\epsilon Z_3\neq 0$ equation
(\ref{chiral}) has the following solutions
\begin{eqnarray}
\epsilon=+1&:&\quad\quad\vec{\chi}_m=(\chi_m,0)\,,\nonumber\\
\epsilon=-1&:&\quad\quad\vec{\chi}_m=(0,\chi_m)\,.
\end{eqnarray}
A non-vanishing spinor breaks $\rSO(3)\times \rSO(2)$ completely.
Equation (\ref{zzs}) reads
\begin{eqnarray}
\overline{Z}_{\bar{S}}&=&-\frac{\epsilon}{8\,\overline{Z}_3}\,\sum_{\bar{m}=1}^2(\overline{\chi}_{\bar{m}})^2\,.
\end{eqnarray}
In the case $\overline{Z}_{\bar{S}}=0$ we need to have either
$\chi_m=0$ or $\chi_1=\pm i\,\chi_2\neq 0$. In the first case the
group $\rSO(2)$ acting on the $m$ index is restored.\par If
$Z_I=0$ a solution of (\ref{ps}) must involve both copies
$\vec{\chi}_m$ and moreover $\vec{\chi}_1=\pm i\,\vec{\chi}_2$.
\subsection{The $L(2,2)$ space}
The global symmetry group in this case is
$\rSO(2)\times\rSO(4)\times \rU(2)$. the spinorial charges $Z_i$
will be denoted as usual by $\vec{\chi}_m\equiv(\chi_{\alpha,m})$,
where $\alpha=1,\dots, 4$ is the spinorial index and $m=1,2$. The
gamma matrices can be chosen of the form
\begin{eqnarray}
{\bf
\Gamma}_2&=&i\,\bfone_2\times\bfone_2\times\bfone_2\,\,;\,\,\,{\bf
\Gamma}_3=\sigma_1\times\sigma_1\times \bfone_2\,\,;\,\,\,{\bf
\Gamma}_4=\sigma_3\times\sigma_1\times \bfone_2\,,\nonumber\\{\bf
\Gamma}_5&=&\bfone_2\times\sigma_3\times \bfone_2\,.
\end{eqnarray}
If $Z_2=-i\epsilon Z_3\neq 0$ and  $Z_4=Z_5=0$, $\rSO(4)$ is
broken to $\rSO(2)$. In this case the solution to (\ref{chiral})
has the form
\begin{eqnarray}
\vec{\chi}_m&=&(\epsilon\,\chi_{1,m},\epsilon\,\chi_{2,m},\chi_{2,m},\chi_{1,m})\,.
\end{eqnarray}
If just one copy of the spinorial charge is non-vanishing,
$\rU(2)$ is broken to $\rU(1)$. Equation (\ref{zzs}) then reads
\begin{eqnarray}
\overline{Z}_{\bar{S}}&=&-\frac{\epsilon}{4\,\overline{Z}_3}\,\sum_{\bar{m}=1}^2\left[(\overline{\chi}_{1,{\bar{m}}})^2+(\overline{\chi}_{2,{\bar{m}}})^2\right]\,.
\end{eqnarray}
If $Z_S=0$ the above equation can be satisfied in three possible
ways: All vanishing spinorial charges, in which case the residual
group is $\rSO(2)\times \rU(2)$;  Just one non vanishing spinorial
charge, in which case the residual group is $\rU(1)$; Two
independent copies of spinorial charges in which case there is no
residual symmetry.\par In the $Z_I=0$ case, the $\rSO(4)$ is
broken by the spinorial charges only. Equation (\ref{ps}), as
opposed to the $q=1$ case, can have non-vanishing solutions with
just one copy of the spinorial charges, in which case $\rU(2)$ is
broken to $\rU(1)$. The possible solutions are
\begin{eqnarray}
\vec{\chi}_m&=&(\chi_{1,m},\,\pm i \chi_{1,m},0,0)\,,\nonumber\\
\vec{\chi}_m&=&(0,0,\chi_{3,m},\,\pm i \chi_{3,m})\,,\nonumber\\
\vec{\chi}_m&=&(\chi_{1,m},\,\pm i \chi_{1,m},\chi_{3,m},\,\mp i \chi_{3,m})\,,\nonumber\\
\vec{\chi}_1&=&\pm i\,\vec{\chi}_2\,.
\end{eqnarray}
If just one copy of the spinorial charges is non vanishing
$\rSO(4)$ is broken to $\rSO(3)_{diag}$.
\section{Conclusions}
In the present paper we have extended the analysis of the
attractor equations to $N=2$ models featuring a generic
homogeneous space. In particular we have found that no deviation
with respect to the symmetric case occurs in the $Z\neq 0$
analysis of non-BPS solutions, namely the Bertotti--Robinson mass
$M_{BR}=2\,|Z|_{extr.}$. We leave for future work the study of the
mass spectra at the critical points and a detailed analysis of
higher curvature corrections which encode possible deviations from
Einstein supergravity.
\section{Acknowledgements}
 Work supported in part by the European
Community's Human Potential Program under contract
MRTN-CT-2004-005104 `Constituents, fundamental forces and
symmetries of the universe', in which R. D'A. and M.T.  are
associated to Torino University. The work of S.F. has been
supported in part by European Community's Human Potential Program
under contract MRTN-CT-2004-005104 `Constituents, fundamental
forces and symmetries of the universe' and the contract MRTN-
CT-2004-503369 `The quest for unification: Theory Confronts
Experiments', in association with INFN Frascati National
Laboratories and by D.O.E. grant DE-FG03-91ER40662, Task C.
\appendix
\section{Some useful properties of special K\"ahler manifolds}\label{ska}
A special K\"ahler manifold $\mathcal{M}$ of complex dimension $n$
is a K\"ahler--Hodge manifold which can be characterized in the
following way (see for example \cite{n2} and references therein).
Let us introduce a $(2n+2)$--dimensional section $V(z,\bar{z})$ of
a symplectic $\rU(1)$ bundle over $\mathcal{M}$. The manifold is
special K\"ahler if $V$ satisfies the following differential
conditions
\begin{eqnarray}
D_r D_s\,V &=& i C_{rsp}\, g^{p\bar p}\bar D_{\bar p}\overline{V}\,,\label{sk1}\\
D_r D_{\bar s}\overline{V} &=& g_{r \bar s } \overline{ V}\,,\label{sk2}\\
D_{s} \overline{V} &=& 0\,, \label{sk3}\end{eqnarray} where the indices $r,s,\dots=1,\dots, n$
are flat indices and the covariant derivative $D_r$ contains the $\rU(1)$ and Levi--Civita
connections. In particular integrability of the above identities require  the Riemann tensor
to have the following form
\begin{eqnarray}
R_p{}^q&=&R_{r\,\bar{s},\,p}{}^q\,V^r\wedge \overline{V}^{\,\,
\bar{s}}\,,\nonumber\\
R_{r\,\bar{s},\,p}{}^q&=&g_{r\bar{s}}\,\delta_p^q+g_{\bar{s}p}\,\delta_r^q-C_{rpk}\,\bar{C}_{\bar{s}}{}^{kp}\,.
\end{eqnarray}
If we write
\begin{eqnarray}
V&=&\left(\matrix{L^\Lambda\cr M_\Sigma}\right)\,\,\,;\,\,\,\,\Lambda\,,\Sigma=0,\dots, n\,,
\end{eqnarray}
the SUSY central charge $Z$ is defined as follows in terms of $V$ and of the quantized
electric and magnetic charges $q_\Lambda,\,p^\Lambda$ as follows
\begin{eqnarray}
Z&=&M_\Lambda\,p^\Lambda-L^\Lambda\,q_\Lambda\,,
\end{eqnarray}
while the matter central charges $Z_r$, in rigid indices, are given by $Z_r=D_r Z$.
\section{Clifford module}\label{spinor}
We use the following $2\,d$--dimensional (reducible)
representation for the $\rSO(2,q+2)$ Clifford algebra:
\begin{eqnarray}
\{\IG_\Lambda,\,\IG_\Sigma\}&=&2\,\eta_{\Lambda\Sigma}\,\,;\,\,\,\,\eta_{\Lambda\Sigma}={\rm diag}(-1,-1,+1,\dots, +1)\,,\nonumber\\
\IG_\Lambda&=&\left(\matrix{{\bf 0}_d&
(\Gamma_\Lambda)_{\alpha\dot{\alpha}}\cr
(\Gamma_\Lambda)^{\dot{\alpha}\alpha}& {\bf
0}_d}\right)\,,\nonumber\\\Lambda,\,\Sigma
&=&0,\dots,q+3;\,\,\,\alpha,\,\dot{\alpha}=1,\dots,d\,.
\end{eqnarray}
The explicit matrix form for $\IG_\Lambda$ that we shall use is,
for $q>0$, the following:
\begin{eqnarray}
\IG_m&=&\gamma_m\times \sigma_2\times
\sigma_2\,\,;\,\,\,m=3,\dots, q+3\,,\nonumber\\
\IG_2&=&\bfone_{\frac{d}{2}}\times \bfone_2\times \sigma_1\,,\nonumber\\
\IG_1&=&\bfone_{\frac{d}{2}}\times i\,\sigma_3\times \sigma_2\,,\nonumber\\
\IG_0&=&\bfone_{\frac{d}{2}}\times i\,\sigma_1\times \sigma_2\,,
\end{eqnarray}
where the $\frac{d}{2}\times \frac{d}{2}$ matrices $\gamma_m$ generate the
$\rSO(q+1)$--Clifford algebra. The charge conjugation matrix is:
\begin{eqnarray}
C^{(-)}&=&\bfone_{\frac{d}{2}}\times
i\,\sigma_2\times\sigma_3=\left(\matrix{C_{\alpha\beta}& {\bf
0}_d\cr {\bf 0}_d&
C^{\dot{\alpha}\dot{\beta}}}\right)\,\,\,;\,\,\,\,C^T=-C\,;\,\,C^2=-\bfone_d\,.
\end{eqnarray}
We also define the (complex)  $\rSO(q+2)$ gamma matrices
\begin{eqnarray}
\hat{\bf \Gamma}_I&=&\left(\matrix{{\bf 0}& {\bf \Gamma}_I\cr
\overline{{\bf \Gamma}}_I&{\bf 0}}\right)\,,
\end{eqnarray}
where the off--diagonal blocks ${\bf \Gamma}_I,\,\overline{{\bf
\Gamma}}_I$ are defined as follows:
\begin{eqnarray}
({\bf \Gamma}_2)_{ij}&=&i\,\delta_{ij}\,\,;\,\,\,({\bf
\Gamma}_m)_{ij}=(\gamma_m)_{ij}\,,\nonumber\\
\overline{\bf \Gamma}_I&=&({\bf \Gamma}_I)^*\label{complexcliff0}
\end{eqnarray}
and satisfy the relation:
\begin{eqnarray}
\{\overline{\bf \Gamma}_I,\,{\bf
\Gamma}_J\}&=&2\,\delta_{IJ}\,.\label{complexcliff}
\end{eqnarray}

\section{Spin connection and the curvature tensor in
components}\label{omegaR} From the Maurer--Cartan equations
(\ref{mc}) we may deduce the structure constants:
\begin{eqnarray}
C_{ab,cI}{}^{dJ}&=&\delta^J_I\,\delta^d_{[a}\,\eta_{b]c}=-\delta^J_I\,\delta^d_{[a}\,\delta_{b]c}\,,\nonumber\\
C_{IJ,aK}{}^{bL}&=&\delta^b_a\,\delta^L_{[I}\,\eta_{J]K}=\delta^b_a\,\delta^L_{[I}\,\delta_{J]K}\,,\nonumber\\
C_{aI,bJ}{}^{cd}&=&-\delta^{cd}_{ab}\,\eta_{IJ}=-\delta^{cd}_{ab}\,\delta_{IJ}\,,\nonumber\\
C_{aI,bJ}{}^{KL}&=&-\delta^{KL}_{IJ}\,\eta_{ab}=\delta^{KL}_{IJ}\,\delta_{ab}\,,\nonumber\\
C_{\Lambda\Sigma,\alpha}{}^\beta
&=&-\frac{1}{4}\,(\Gamma_{\Lambda\Sigma})_\alpha{}^\beta\,,\nonumber\\
C_{\alpha\beta}{}^\bullet&=&2\,C_{\alpha\beta}\,\,\,;\,\,\,
C_{0\alpha}{}^{\beta}=\delta_\alpha^\beta\,\,\,;\,\,\, C_{0\bullet}{}^{\bullet}=2\,.
\end{eqnarray}
The non--zero constant components $\tilde{\omega}_{A,B}{}^C$ of
the spin connection are:
\begin{eqnarray}
\tilde{\omega}_{\bullet,\alpha}{}^\beta&=&\frac{1}{2}\,C_{\beta^\prime
\alpha}{}^\bullet\,g^{\beta\beta^\prime}\,g_{\bullet\bullet}=-C_{\alpha\beta}\,,\nonumber\\
\tilde{\omega}_{\alpha,\bullet}{}^\beta&=&\frac{1}{2}\,C_{\beta^\prime
\alpha}{}^\bullet\,g^{\beta\beta^\prime}\,g_{\bullet\bullet}=-C_{\alpha\beta}\,,\nonumber\\
\tilde{\omega}_{\alpha\beta}{}^\bullet&=&C_{\alpha\beta}\,,\nonumber\\
\tilde{\omega}_{ab,\alpha}{}^\beta&=&-\frac{1}{4}\,(\Gamma_{ab})_\alpha{}^\beta\,\,;\,\,\,\tilde{\omega}_{IJ,\alpha}{}^\beta=-\frac{1}{4}\,(\Gamma_{IJ})_\alpha{}^\beta\,,\nonumber\\
\tilde{\omega}_{\bullet,0}{}^\bullet&=&-2\,\,;\,\,\,\tilde{\omega}_{\bullet,\bullet}{}^0=2\,g^{00}\,g_{\bullet\bullet}=2\,,\nonumber\\
\tilde{\omega}_{\alpha,0}{}^\beta&=&-\delta_\alpha^\beta\,\,;\,\,\,\tilde{\omega}_{\alpha,\beta}{}^=g_{\beta\alpha}\,g^{00}=\delta_{\alpha\beta}\,,\nonumber\\
\tilde{\omega}_{\alpha,aI}{}^\beta&=&\frac{1}{4}\,(\Gamma_{aI})_\alpha{}^\beta\,\,;\,\,\,\tilde{\omega}_{\alpha,\beta}{}^{aI}=-\frac{1}{4}\,(\Gamma_{a^\prime
I^\prime})_\alpha{}^{\beta^\prime}\,g^{a^\prime I^\prime
aI}\,g_{\beta\beta^\prime}=-2\,(\Gamma_{aI})_\alpha{}^\beta\,.\label{omcomponents}
\end{eqnarray}
For completeness, let us give the explicit form of the components
of the $H$--connection $\Omega^{(H)}$ in (\ref{omomt}), obtained
by solving the second of eqs. (\ref{mc})
\begin{eqnarray}
\Omega^{(H)}{}_{aI,}{}^c{}_b&=&\left(\frac{2}{q+2}\right)\,(V_{bJ}{}^ u\,V_{aI}{}^ v\,\partial_{[ u}V_{ v]}{}^{cJ}+V_{aI}{}^ u\,V^{cJ\, v}\,\partial_{[ u}V_{ v]\,bJ}-V^{cJ\, u}\,V_{bJ}{}^ v\,\partial_{[ u}V_{ v]\,aI})\,,\nonumber\\
\Omega^{(H)}{}_{aI,}{}^K{}_J&=&V_{bJ}{}^ u\,V_{aI}{}^
v\,\partial_{[ u}V_{ v]}{}^{bK}+V_{aI}{}^ u\,V^{bK\,
v}\,\partial_{[ u}V_{ v]\,bJ}-V^{bK\, u}\,V_{bJ}{}^ v\,\partial_{[
u}V_{ v]\,aI}\,,\label{omH}
\end{eqnarray}
$u,\,v=1,\dots, 2n $ are curved indices on the scalar manifold and
$V_{aI}{}^ u$ denotes the inverse vielbein:
\begin{eqnarray}
V_{aI}{}^ u\,V_ u{}^{bJ}&=&\delta_a^b\,\delta_I^J\,.
\end{eqnarray}
 From eq. (\ref{rcomp}) we deduce the
components of the Riemann tensor:
\begin{eqnarray}
R_{aI\,
bJ,cK}{}^{dL}&=&-\frac{1}{2}\,\delta^d_{[a}\,\eta_{b]c}\,\eta_{IJ}\,\delta^{K}_L-\frac{1}{2}\,\delta^L_{[I}\,\eta_{J]K}\,\eta_{ab}\,\delta^{d}_c=\nonumber\\
&=&\frac{1}{2}\,\delta^d_{[a}\,\delta_{b]c}\,\delta_{IJ}\,\delta^{K}_L+\frac{1}{2}\,\delta^L_{[I}\,\delta_{J]K}\,\delta_{ab}\,\delta^{d}_c\nonumber\\
R_{\alpha\,\beta,aI}{}^{bJ}&=&\frac{1}{8}\,(\Gamma_{ab^\prime})_\alpha{}^{\beta^\prime}\,\eta_{IJ^\prime}\,g^{b^\prime
J^\prime}\,g_{\beta\beta^\prime}
+\frac{1}{8}\,(\Gamma_{IJ^\prime})_\alpha{}^{\beta^\prime}\,\eta_{ab^\prime}\,g^{b^\prime
J^\prime}\,g_{\beta\beta^\prime}=\nonumber\\&=&(\Gamma_{ab})_\alpha{}^{\beta}\,
\delta_{IJ}
-(\Gamma_{IJ})_\alpha{}^{\beta}\,\delta_{ab}\,,\nonumber\\
R_{aI\,\alpha,bJ}{}^\beta &=&-
\frac{1}{16}\,(\Gamma_{aI})_\alpha{}^\gamma\,(\Gamma_{bJ})_\gamma{}^\beta\,,\nonumber\\
R_{\alpha\,\beta,0}{}^\bullet &=&-2\,C_{\alpha\beta}\,,\nonumber\\
R_{\alpha\,\bullet,aI}{}^{\beta}&=&-\frac{1}{4}\,(\Gamma_{aI})_\alpha{}^\gamma\,C_{\gamma\beta^\prime}\,g^{\beta^\prime\beta}\,g_{\bullet\bullet}=
-\frac{1}{4}\,(\Gamma_{aI})_\alpha{}^\gamma\,C_{\gamma\beta}\,,\nonumber\\
R_{\alpha\,\bullet,\,0}{}^\beta&=&-C_{\alpha\beta^\prime}\,g^{\beta^\prime\beta}=-C_{\alpha\beta}\,,\nonumber\\
R_{aI\,\alpha,\,0}{}^\beta&=&\frac{1}{4}\,(\Gamma_{aI})_\alpha{}^\beta\,,\nonumber\\
R_{0\,\alpha,\,aI}{}^\beta&=&\frac{1}{4}\,(\Gamma_{aI})_\alpha{}^\beta\,,\nonumber\\
R_{0\,\alpha,\,0}{}^\beta&=&-\delta_\alpha^\beta\,,\nonumber\\
R_{0\,\bullet,\,0}{}^\bullet&=&-4\,,\nonumber\\
R_{aI\,bJ,\,\alpha}{}^\beta
&=&\frac{1}{8}\,(\Gamma_{ab})_\alpha{}^{\beta}\, \delta_{IJ}
-\frac{1}{8}\,(\Gamma_{IJ})_\alpha{}^{\beta}\,\delta_{ab}\,,\nonumber\\
R_{aI\,\alpha,\,\beta}{}^\bullet&=&-\frac{1}{4}\,(\Gamma_{aI})_\alpha{}^\gamma\,C_{\gamma\beta}\,,\nonumber\\
R_{0\,\bullet,\,\alpha}{}^\beta&=&-2\,C_{\alpha\gamma}\,g^{\gamma\beta}=-2\,C_{\alpha\beta}\,,\nonumber\\
R_{0\,\alpha,\,\beta}{}^\bullet
&=&C_{\alpha\beta}\,,\nonumber\\
R_{\alpha\,\beta,\,\gamma}{}^\delta&=&-2\,C_{\gamma\delta^\prime}\,C_{\alpha\beta}\,g^{\delta^\prime\delta}+2\,C_{\gamma[\alpha}\,C_{\beta]\delta^\prime}\,g^{\delta^\prime\delta}-
2\,g_{\gamma[\alpha}\,g_{\beta]\delta^\prime}\,g^{\delta^\prime\delta}-\frac{1}{8}\,(\Gamma_{a^\prime
I^\prime})_{[\alpha}{}^{\gamma^\prime}\,(\Gamma_{aI})_{\beta]}{}^\delta\,g_{\gamma^\prime\gamma}\,g^{a^\prime
I^\prime,aI}=\nonumber\\
&=&-2\,C_{\gamma\delta}\,C_{\alpha\beta}+2\,C_{\gamma[\alpha}\,C_{\beta]\delta}-
2\,g_{\gamma[\alpha}\,g_{\beta]\delta}-(\Gamma_{a
I})_{[\alpha}{}^{\gamma}\,(\Gamma_{aI})_{\beta]}{}^\delta\,\nonumber\\
R_{\alpha\,\bullet,\,\beta}{}^\bullet
&=&-g_{\alpha\beta}=-\delta_{\alpha\beta}\,.\label{Rcomponent}
\end{eqnarray}
\section{The $\rSO(1,q+1)$--covariant rigid basis}
With the exception of the \emph{minimal coupling} case
$L(-2,\,P)=\rU(1,1+P)/\rU(1)\times \rU(1+P)$, all the other
homogeneous special K\"ahler manifolds have a five dimensional
``parent'' manifold which is a \emph{real special} manifold. This
manifold always admits an $\rSO(1,1+q)$ group of isometries,
which, in most of the cases, can be extended to global symmetry
group of the whole theory. We may keep track of the five
dimensional origin of a special K\"ahler manifold, by choosing a
basis for the tangent space of the four dimensional manifold which
is $\rSO(1,1+q)$--covariant. To this end we decompose the
$\mathfrak{so}(2,2+q)$ algebra as follows
\begin{eqnarray}
\mathfrak{so}(2,2+q)&=&\mathfrak{o}(1,1)_1\oplus\mathfrak{so}(1,1+q)\oplus
{\bf (2+q)}_{\frac{1}{2}}\oplus {\bf (2+q)}_{-\frac{1}{2}}\,,
\end{eqnarray}
where the subscripts refer to the $\mathfrak{o}(1,1)_1$--grading,
and choose as a basis for the tangent space of the
$\rSO(2,2+q)/\rSO(2)\times\rSO(2+q)$ sub-manifold the following
isometry generators
\begin{eqnarray}
T_I &=&\{T_{12},\,T_{0m}\}\,\,\,;\,\,\,\,I=2,\dots,
q+3\,\,\,;\,\,\,\,m=3,\dots, q+3\,,\nonumber\\
T_{+I}&=&\{T_{10}+T_{20},\,T_{1m}+T_{2m}\}\,,
\end{eqnarray}
where $T_{12}$ is the $\mathfrak{o}(1,1)_1$ generator and is
parametrized by the radial modulus of the internal circle in the
reduction from five to four dimensions, $T_{1m}$ generate the
coset $\rSO(1,1+q)/\rSO(1+q)$ and $T_{+I}$ are the generators of
the $(q+2)$--dimensional abelian subalgebra (translations) ${\bf
(2+q)}_{\frac{1}{2}}$. The new basis $\{T^\prime_A\}$ for the
tangent space of the total manifold is now
\begin{eqnarray}
\{T^\prime_A\}&=&\{h_0,\,T_{I},\,T_{+I},\,T_\alpha,\,T_\bullet\}\,,
\end{eqnarray}
and we shall denote by
$\{V^{A}\}=\{V^0,\,V^{I},\,V^{+I},\,V^\alpha,\,V^\bullet\}$ the
dual vielbein basis. We then define the complex vielbein as
follows
\begin{eqnarray}
V^S&=&\frac{1}{\sqrt{2}}\,(V^\bullet-i\,V^0)\,\,;\,\,\,V^I=\frac{1}{\sqrt{2}}\,(V^{+I}+i\,V^{I})\,,\nonumber\\
V^i&=&\frac{1}{\sqrt{2}}\,(V^{i_1}+i\,V^{i_2})\,\,\,,\,\,\,\,\alpha=(i_1,\,i_2),\,\,\,i_1,i_2=1,\dots,\frac{d}{2}\,,
\end{eqnarray}
namely the abelian subalgebra ${\bf (2+q)}_{\frac{1}{2}}$ is
parametrized by the real parts of the complex scalars which span
the submanifold $\rSO(2,2+q)/\rSO(2)\times\rSO(2+q)$.
 If we write the curvature in this new basis we find the following
 non-vanishing components
\begin{eqnarray}
R_{r\,\bar{s},\,p}{}^q&=&g_{r\bar{s}}\,\delta_p^q+g_{\bar{s}p}\,\delta_r^q-
C^{\,\,\prime}_{rpk}\,\bar{C}^{\,\,\prime}_{\bar{s}}{}^{kq}\,,
\end{eqnarray}
where the new symmetric tensor $C^{\,\,\prime}_{rsp}$ has the form
\begin{eqnarray}
C^{\,\,\prime}_{SIJ}&=&\frac{1}{8}\,\eta_{IJ}\,\,\,;\,\,\,\,C^{\,\,\prime}_{Iij}=\frac{1}{4}\,({\bf
\Gamma}^\prime_I)_{ij}\,,
\end{eqnarray}
$\eta_{IJ}=\mbox{diag}(-1,+1,\dots,+1)$ being the
$\rSO(1,1+q)$--invariant tensor of the ${\bf q+2}$ representation
and ${\bf \Gamma}_I^\prime,\,\overline{{\bf \Gamma}}_I^\prime$,
defined as
\begin{eqnarray}
({\bf \Gamma}^{\,\prime}_2)_{ij}&=&-\delta_{ij}\,\,;\,\,\,({\bf
\Gamma}^{\,\prime}_m)_{ij}=(\gamma_m)_{ij}\,,\nonumber\\
(\overline{{\bf
\Gamma}}^{\,\prime}_2)_{ij}&=&\delta_{ij}\,\,;\,\,\,(\overline{{\bf
\Gamma}}^{\,\prime}_m)_{ij}=(\gamma_m)_{ij}\,,
\end{eqnarray}
are blocks of $\rSO(1,1+q)$ gamma--matrices
\begin{eqnarray}
\hat{\bf \Gamma}^\prime_I&=&\left(\matrix{{\bf 0}& {\bf
\Gamma}^\prime_I\cr \overline{{\bf \Gamma}}^\prime_I&{\bf
0}}\right)\,\,\,\,;\,\,\, \{\hat{\bf
\Gamma}^{\,\prime}_I,\,\hat{\bf
\Gamma}^{\,\prime}_J\}=2\,\eta_{IJ}\,.
\end{eqnarray}
Note that the curvature and the $C^{\,\,\prime}$ tensors are now
$\rSO(1,1+q)$--invariant and that the change from the
$\rSO(2+q)$--covariant vielbein basis to the
$\rSO(1,1+q)$--covariant one amounts to the action of the
following $\rU(n)$ transformation:
\begin{eqnarray}
V^{I=2}&\rightarrow & i\,V^{I=2}\,\,\,,\,\,\,\,V^{I\neq
2}\rightarrow  V^{I\neq 2}\,\,\,,\,\,\,\,V^{A\neq I}\rightarrow
V^{A\neq I}\,.
\end{eqnarray}


\begin{thebibliography}{99}
\bibitem{attractor}
 S.~Ferrara, R.~Kallosh and A.~Strominger,
  ``N=2 extremal black holes,''
  Phys.\ Rev.\ D {\bf 52}, 5412 (1995)
  [arXiv:hep-th/9508072];
   A.~Strominger,
  ``Macroscopic Entropy of $N=2$ Extremal Black Holes,''
  Phys.\ Lett.\ B {\bf 383} (1996) 39
  [arXiv:hep-th/9602111];
   S.~Ferrara and R.~Kallosh,
  ``Supersymmetry and Attractors,''
  Phys.\ Rev.\ D {\bf 54}, 1514 (1996)
  [arXiv:hep-th/9602136];
 S.~Ferrara and R.~Kallosh,
 ``Universality of Supersymmetric Attractors,''
  Phys.\ Rev.\ D {\bf 54} (1996) 1525
  [arXiv:hep-th/9603090].

  \bibitem{fgk} S.~Ferrara, G.~W.~Gibbons and R.~Kallosh,
  ``Black holes and critical points in moduli space,''
  Nucl.\ Phys.\ B {\bf 500}, 75 (1997)
  [arXiv:hep-th/9702103].
  \bibitem{fk} S.~Ferrara and R.~Kallosh,
  ``On N = 8 attractors,''
  Phys.\ Rev.\ D {\bf 73}, 125005 (2006)
  [arXiv:hep-th/0603247].

\bibitem{gijt}  K.~Goldstein, N.~Iizuka, R.~P.~Jena and S.~P.~Trivedi,
  ``Non-supersymmetric attractors,''
  Phys.\ Rev.\ D {\bf 72} (2005) 124021
  [arXiv:hep-th/0507096].

  \bibitem{k} R.~Kallosh,
  ``New attractors,''
  JHEP {\bf 0512} (2005) 022
  [arXiv:hep-th/0510024].

\bibitem{tt0}
 P.~K.~Tripathy and S.~P.~Trivedi,
  ``Non-supersymmetric attractors in string theory,''
  JHEP {\bf 0603} (2006) 022
  [arXiv:hep-th/0511117].

\bibitem{g} A.~Giryavets,
  ``New attractors and area codes,''
  JHEP {\bf 0603} (2006) 020
  [arXiv:hep-th/0511215].

  \bibitem{kss}
  R.~Kallosh, N.~Sivanandam and M.~Soroush,
  ``The non-BPS black-hole attractor equation,''
  JHEP {\bf 0603} (2006) 060
  [arXiv:hep-th/0602005].

\bibitem{cdwkm}
  G.~Lopes Cardoso, B.~de Wit and T.~Mohaupt,
  ``Corrections to macroscopic supersymmetric black-hole entropy,''
  Phys.\ Lett.\ B {\bf 451} (1999) 309
  [arXiv:hep-th/9812082];
G.~Lopes Cardoso, B.~de Wit, J.~Kappeli and T.~Mohaupt,
  ``Stationary BPS solutions in N = 2 supergravity with R**2 interactions,''
  JHEP {\bf 0012} (2000) 019
  [arXiv:hep-th/0009234];
G.~Lopes Cardoso, B.~de Wit, J.~Kappeli and T.~Mohaupt,
  ``Black hole partition functions and duality,''
  JHEP {\bf 0603} (2006) 074
  [arXiv:hep-th/0601108].
  \bibitem{sv} A.~Strominger and C.~Vafa,
  ``Microscopic Origin of the Bekenstein-Hawking Entropy,''
  Phys.\ Lett.\ B {\bf 379}, 99 (1996)
  [arXiv:hep-th/9601029].
  \bibitem{msw} J.~M.~Maldacena, A.~Strominger and E.~Witten,
  ``Black hole entropy in M-theory,''
  JHEP {\bf 9712}, 002 (1997)
  [arXiv:hep-th/9711053].
 \bibitem{osv} H.~Ooguri, A.~Strominger and C.~Vafa,
  ``Black hole attractors and the topological string,''
  Phys.\ Rev.\ D {\bf 70} (2004) 106007
  [arXiv:hep-th/0405146].
  \bibitem{d} A.~Dabholkar,
  ``Exact counting of black hole microstates,''
  Phys.\ Rev.\ Lett.\  {\bf 94}, 241301 (2005)
  [arXiv:hep-th/0409148].
  \bibitem{ddmp}A.~Dabholkar, F.~Denef, G.~W.~Moore and B.~Pioline,
  ``Precision counting of small black holes,''
  JHEP {\bf 0510}, 096 (2005)
  [arXiv:hep-th/0507014].
  \bibitem{s}  A.~Sen,
  ``Black hole entropy function and the attractor mechanism in higher
  derivative gravity,''
  JHEP {\bf 0509}, 038 (2005)
  [arXiv:hep-th/0506177].
  \bibitem{dst}A.~Dabholkar, A.~Sen and S.~Trivedi,
  ``Black hole microstates and attractor without supersymmetry,''
  arXiv:hep-th/0611143;
  \bibitem{p} B.~Pioline,
  ``Lectures on on black holes, topological strings and quantum attractors,''
  Class.\ Quant.\ Grav.\  {\bf 23} (2006) S981
  [arXiv:hep-th/0607227].
\bibitem{bfgm} S.~Bellucci, S.~Ferrara, M.~Gunaydin and A.~Marrani,
  ``Charge orbits of symmetric special geometries and attractors,''
  Int.\ J.\ Mod.\ Phys.\ A {\bf 21}, 5043 (2006)
  [arXiv:hep-th/0606209].
\bibitem{a}
D.~V. Alekseevsky, ``Classification of quaternionic spaces with a
  transitive solvable group of motions'', Math.\ USSR Izvestija {\bf 9} (1975)
297--339.
\bibitem{dwvvp}
 B.~de~Wit, F.~Vanderseypen  and A.~Van~Proeyen,
``Symmetry structure of
  special geometries'', Nucl. Phys. {\bf B400} (1993) 463--524,
hep-th/9210068;
\bibitem{c}V.~Cort{\'e}s, ``Alekseevskian spaces'', Diff. Geom. Appl.
{\bf 6} (1996) 129--168.
\bibitem{tt} P.~K.~Tripathy and S.~P.~Trivedi,
  ``Compactification with flux on K3 and tori,''
  JHEP {\bf 0303} (2003) 028
  [arXiv:hep-th/0301139].
\bibitem{aft}  L.~Andrianopoli, R.~D'Auria, S.~Ferrara and M.~A.~Lledo,
  ``4-D gauged supergravity analysis of type IIB vacua on K3 x T**2/Z(2),''
  JHEP {\bf 0303} (2003) 044
  [arXiv:hep-th/0302174].
\bibitem{adft} C.~Angelantonj, R.~D'Auria, S.~Ferrara and M.~Trigiante,
  ``K3 x T**2/Z(2) orientifolds with fluxes, open string moduli and  critical
  points,''
  Phys.\ Lett.\ B {\bf 583} (2004) 331
  [arXiv:hep-th/0312019].
\bibitem{n2}
 L.~Andrianopoli, M.~Bertolini, A.~Ceresole, R.~D'Auria, S.~Ferrara, P.~Fre and T.~Magri,
  ``N = 2 supergravity and N = 2 super Yang-Mills theory on general scalar
  manifolds: Symplectic covariance, gaugings and the momentum map,''
  J.\ Geom.\ Phys.\  {\bf 23} (1997) 111
  [arXiv:hep-th/9605032].
\end{thebibliography}
\end{document}